\documentclass[twocolumn]{aa}
\usepackage{graphicx}
\usepackage{txfonts}
\usepackage{longtable}
\begin{document}
   \title{Footpoint versus loop-top hard X--ray emission sources\\ in solar flares}

   \author{M. Tomczak
          \and
           T. Ciborski}

   \offprints{M. Tomczak}

   \institute{Astronomical Institute, University of Wroc{\l }aw,
              ul. Kopernika 11, PL-51--622 Wroc{\l }aw, Poland,
              \email{tomczak@astro.uni.wroc.pl}
             }

   \date{Received 26 June 2006 / Accepted 5 September 2006}

   \abstract
   {}
   {The hard X-ray flux ratio $\Re$ of the footpoint sources to the
   loop-top source has been used to investigate non-thermal
   electron trapping and precipitation in solar flares.}
   {Considering the mission-long Yohkoh Hard X--ray Telescope database,
   from which we selected 117 flares, we investigated a dependence
   of the ratio $\Re$ on flare loop parameters like height $h$ and
   column depth $N$. We used non-thermal electron beams as a diagnostic
   tool for magnetic convergence.}
   {The ratio $\Re$ decreases with $h$ which we interpret as
   an effect of converging field geometry. Two branches seen in the $\Re-h$
   diagram suggest that in the solar corona two kinds of magnetic
   loops can exist: a more-converged ones that are more frequent (above 80\%) and
   less-converged loops that are less frequent (below 20\%). A lack of correlation
   between the ratio $\Re$ and $N$ can be due to a more complex
   configuration of investigated events than seen in soft X--rays.}
   {Obtained values of the magnetic mirror ratio are consistent with previous
   works and suggest a strongly nonpotential configuration. Further
   investigation including RHESSI data is needed to verify our results.}


   \maketitle
%

\section{Introduction}
\label{intr}

Images taken by the Hard X--ray Telescope (HXT) onboard {\sl Yohkoh}
show different spatial distributions of hard X--ray emission in
solar flares. Classifications of hard X--ray emission sources
observed in solar flares have been reported (e.g. Kosugi
\cite{kosugi}, Masuda \cite{masuda2}, Aschwanden \cite{aschwpodr}).
However, an unification is possible in which hard X--ray emission
sources are divided into two classes: coronal and chromospheric.
Members of the first class usually occur within or above the top of
the magnetic coronal loop clearly seen in soft X--rays, thus we call
them {\sl loop-top} or {\sl above-the-loop-top} sources. Members of
the second class occur at the entrance of the same magnetic coronal
loops into the chromosphere, thus we call them {\sl footpoint}
sources. For the more complicated morphologies, interacting loops or
an arcade of loops, such a division still works, however, more hard
X--ray sources can be seen at times.

The majority of flares show the presence of hard X--ray emission
sources of both classes: loop-top source and footpoint ones.
However, their basic properties are different. The footpoint sources
are stronger than the loop-top ones during the hard X--ray bursts of
the impulsive phase. Footpoint sources grow with higher-energy
photons. At the end of the impulsive phase the footpoint sources
fade and the loop-top source becomes brighter but its radiation only
rarely exceeds 30~keV.

Energy spectra of hard X--ray photons emitted during the impulsive
phase of flares have a non-thermal shape. They can be described by a
single or a double power-law function. Usually the energy spectra of
footpoint sources are flatter than the spectra of the loop-top
sources i.e. the index $\gamma$ of the function $I(E) = A
E^{-\gamma}$ is smaller for the footpoint sources than for the
loop-top ones. Spectra of loop-top sources adopt a quasi-thermal
shape with time i.e. an increase of the index $\gamma$ with the
energy of photons is seen.

Both classes of hard X--ray emission sources have different
mechanism of radiation. The loop-top sources are connected directly
to the process of energy release in flares. Due to magnetic
reconnection some electrons are accelerated up to high energies and
they immediately radiate their energy excess in the loop-top region
via thin-target or thermal bremsstrahlung. In some models the area
occupied by the loop-top source defines the reconnection site
(Jakimiec et al. \cite{jakimiec}, Petrosian \& Donaghy
\cite{petrosian}). In other models, the reconnection X--point occurs
higher in the corona and the loop-top source is a consequence of
interaction between a downward reconnection outflow and plasma
filling a soft X--ray flare loop (Tsuneta \cite{tsuneta}, Shibata
\cite{shibata}). The motion of the magnetic field lines in the cusp
structure below the reconnection site and above the underlying flare
loops, which we call the collapsing magnetic trap, also can be an
efficient accelerator of electrons (Somov \& Kosugi \cite{somov},
Jakimiec \cite{jj}, Karlick\'y \& Kosugi \cite{karl1}, Karlick\'y \&
B\'arta \cite{karl3}). The footpoint sources are regions in which
accelerated electrons are precipitated into the chromosphere and
radiate via thick-target bremsstrahlung (Brown \cite{brown}).

A simultaneous investigation of the loop-top and footpoint hard
X--ray emission sources offers a chance to better understand the
main physical processes occurring during the impulsive phase in
solar flares. The first attempt of such a kind of analysis was made
by Masuda (\cite{masuda}) who compared some characteristics of
loop-top and footpoint sources for 10 single-loop flares observed by
{\sl Yohkoh} between 1991 and 1993. He reported the presence of hard
X--ray emission sources of both classes in the majority of
investigated flares. He found that during the impulsive phase the
loop-top source mimics a variability of footpoint sources. He also
pointed out that the coronal sources shifted above the loop-top are
systematically more energetic (with a lower index $\gamma$) than the
sources that are co-spatial with the loop-top.

More comprehensive analysis was performed by Petrosian et al.
(\cite{petrosian2}). They selected and analyzed HXT images of 18
flares from 1991--1998 and included multiple-loop events. The
authors conclude that the loop-top and footpoint hard X--ray
emission sources are a common characteristic of the impulsive phase
of solar flares. They found an obvious correlation between loop-top
and footpoint hard X--ray fluxes over more than two decades of flux,
and obtained that the footpoint sources are systematically stronger
and more energetic (with a lower index $\gamma$) than loop-top ones.

In this paper we extend the previous studies. We discuss the
relation between the loop-top and footpoint hard X--ray emission
sources in solar flares for a larger number of events. Our database
includes flares observed over the entire {\sl Yohkoh} mission
(1991--2001) and we modified the selection criteria.

The paper is organized as follows. In Section~\ref{selection} a
short description of scientific instruments is given as are the
selection criteria. In Section~\ref{ratio} we define the flux ratio
$\Re$ of the footpoint and loop-top hard X--ray emission sources in
investigated flares and present its dependence on other parameters
such height of the flare loop and column density along the flare
loop. In Section~\ref{disc} we discuss the implications of our
results for the magnetic field configuration in solar flares. In
Section~\ref{concl} we present a summary and our conclusions.

\section{Scientific instruments and data selection}
\label{selection}

We used data from two imaging instruments onboard {\sl Yohkoh}: the
HXT and the Soft X--ray Telescope (SXT). The HXT (Kosugi et al.
\cite{kosugi_rb}) is a Fourier synthesis imager observing the whole
Sun. It consisted of 64 independent subcollimators which measured
spatially modulated intensities in four energy bands (L: 14--23~keV,
M1: 23--33~keV, M2: 33--53~keV, and H:53--93~keV). During the flare
the intensities were integrated, in each energy band, over 0.5~s.
Some reconstruction routines are available that allow us to obtain
hard X--ray images with an angular resolution of up to 5 arcsec. We
used the Maximum Entropy Method developed for HXT data by Sakao
(\cite{sakao}). This method works very efficiently since the
in-orbit calibration of the HXT response function was performed
(Sato et al. ~\cite{sato}).

The SXT (Tsuneta et al. \cite{tsuneta_rb}) is a grazing-incidence
telescope sensitive to 0.4--4.2~keV soft X--rays, with a CCD
detector and filters to provide wavelength discrimination. During
flare mode the SXT usually recorded frames of 64~$\times$~64 pixels.
Images were taken sequentially with different filters every 2
seconds and their spatial resolution was 2.45 arcsec$^2$. In our
analysis we used two filters: a 119 $\mu$m beryllium filter (Be119)
and a 11.6 $\mu$m aluminium filter (Al12). The signal ratio of these
filters allows us to estimate the temperature and emission measure
of the soft X--ray flare plasma (Hara et al. \cite{hara}).

Masuda (\cite{masuda}) used two criteria for flare selection: a
heliocentric longitude greater than 80$^{\circ}$, and peak count
rate in the M2 band greater than 10 counts per second per
subcollimator. The first criterion ensures maximum angular
separation between loop-top and footpoint sources. The second
criterion ensures that at least one image can be obtained at
energies where the thermal contribution should be negligible. The
same criteria was used by Petrosian et al. (\cite{petrosian2}).

In our analysis we slightly modified these criteria because we found
out them too strict. We qualified flares that occurred at
heliocentric longitudes greater than 65$^{\circ}$, instead of
80$^{\circ}$, and had a peak count rate in the M1 band greater than
10 counts per second per subcollimator, instead of the M2 band. For
flares between 65$^{\circ}$ and 80$^{\circ}$ a separation between
loop-top and footpoint sources is still possible. We limited our
analysis presented in Section~\ref{ratio} to the moment of peak
count rate in the band M1, instead of the time interval over the
entire impulsive duration of flares which has been used in previous
works. At that moment, emission of non-thermal electrons strongly
dominates the thermal contribution, which is proved by almost the
same moment of peak count rate in the bands M1, M2, and H for the
majority of flares.

In the catalog including flares observed over the entire {\sl
Yohkoh} mission
(1991--2001)\footnote{http://solar.physics.montana.edu/sato/shxtdbase.html}
we found that our criteria were fulfilled by 198 out of 3071 events.
Among these 198 flares we excluded some events due to: a lack of SXT
images, poor quality of HXT images, and location behind the solar
limb where we can observe loop-top sources only. We then considered
117 flares presented in Table~\ref{113}. Such a large number of
analyzed events offers the chance to statistically model the
relation between loop-top and footpoint hard X--ray emission sources
in solar flares. Our list contains almost all the flares analyzed
previously by Masuda (\cite{masuda}) and Petrosian et al.
(\cite{petrosian2}). Excluded are the flare of April 23, 1998 which
occurred 13$^{\circ}$ behind the solar limb (Sato \cite{sato2}) and
the flares of May 8 and 9, 1998 for which there were no SXT images
for the investigated time period.

\section{The flux ratio $\Re$ of hard X--ray sources}
\label{ratio}

\begin{figure*}
\resizebox{\hsize}{!}{\includegraphics{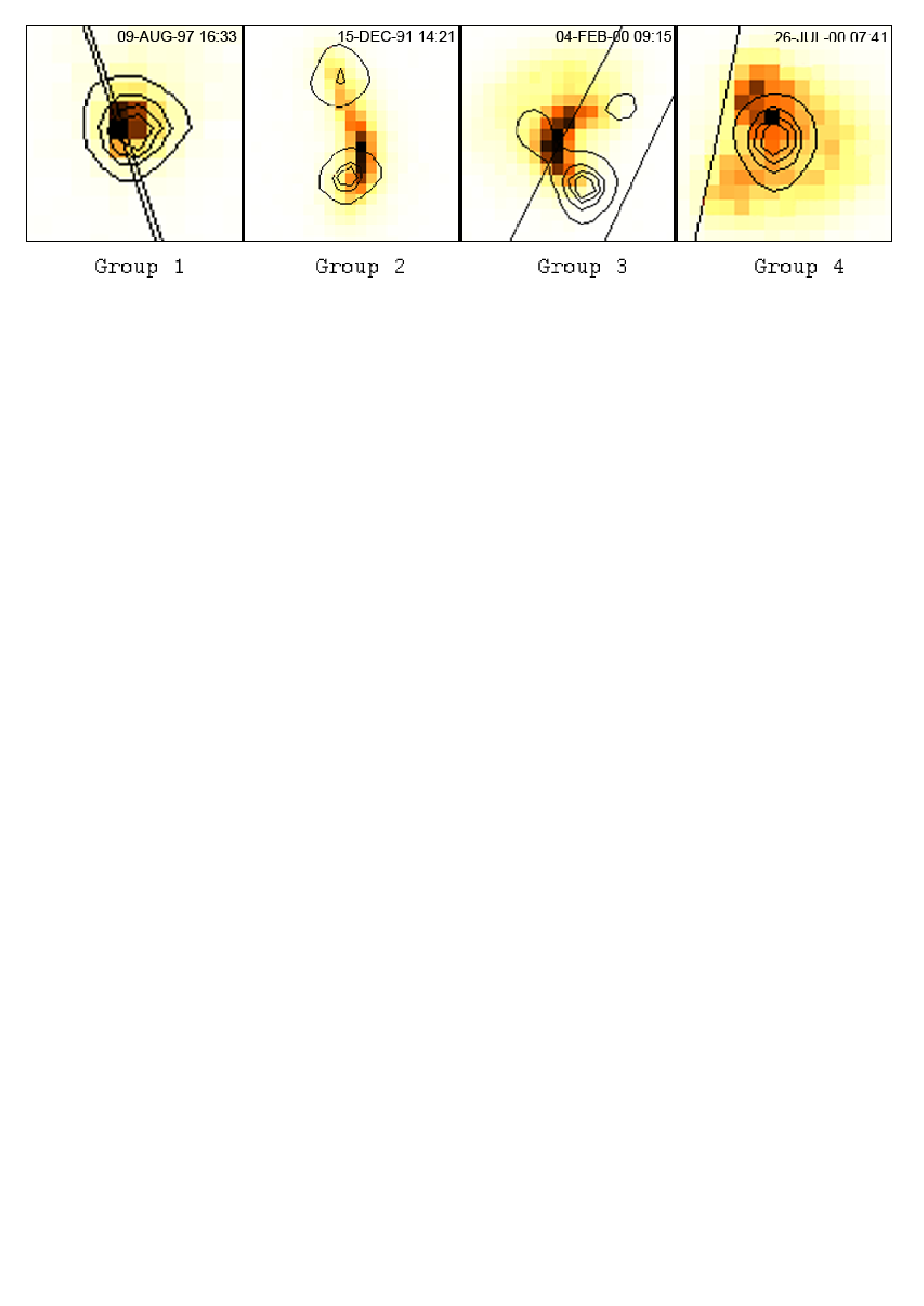}} \hfill
\caption{Examples of groups introduced in the text. The contours and
the half-tones show the HXT/M1 and SXT/Be119 images, respectively.
The pixel size is 2.45\,arcsec$^2$. North is at the top, west -- to
the right.} \label{4g}
\end{figure*}

For each flare from the list we computed one image at the peak count
rate in the M1 band. To obtain an image of good quality we need
about 200 counts per subcollimator, therefore we accumulated signal
within a time interval of between 0.5 and 11~s (see column (7) in
Table~\ref{113}), depending on the total number of counts in the
band M1. Next, we carefully defined the shape of the individual
emission sources and combined the signal from all pixels inside each
source separately. Due to the limited dynamic range of the HXT and
the image reconstruction process, estimated to be about one decade
(Sakao \cite{sakao}), we considered as background the pixels having
a signal below 10\% of the brightest one. Sometimes emission sources
in the M1 band were not well resolved, thus we calculated images in
higher energy bands, if available, to define the border between the
sources more precisely.

Keeping in mind that the loop-top and footpoint hard X--ray emission
sources are a common characteristic of the impulsive phase of solar
flares (e.g. Tomczak \cite{tomczak-occulted}), we recognized four
groups of events (see column (8) in Table~\ref{113}):
\begin{itemize}
\item The group 1 (35 events) for which the HXT could not resolve individual sources.
\item The group 2 (37 events) for which the HXT detected only footpoint sources.
\item The group 3 (40 events) for which the HXT detected footpoint and loop-top
sources.
\item The group 4 (5 events) for which the HXT detected only loop-top sources.
\end{itemize}
Examples of events from each group are presented in Fig.~\ref{4g}.
Events from group 1 are generally too small for the imaging
capability of the HXT, thus we excluded them from further analysis.
The presence of groups 2 and 4 is likely caused by the limited
dynamic range of the HXT. For events from group 2, footpoint sources
are brighter than the loop-top source by more than 10 times,
opposite to events from group 4 for which the loop-top source is
brighter than footpoint sources by more than 10 times. In both cases
the fainter sources cannot be resolved from the background but we
expect that they exist.

Following Petrosian et al. (\cite{petrosian2}) we define a parameter $\Re$ which is
a ratio of the sum of the counts of the footpoint sources to that of the loop-top
sources.
\begin{equation}
\Re = F_{FP}/F_{LT}
\end{equation}
As a consequence, events from group 2 have values of $\Re$ greater
than 10, events from group 4 have values of $\Re$ less than 0.1, and
events from group 3 have values of $\Re$ between 10 and 0.1 (see
column (9) in Table~\ref{113}).

\begin{figure*}
\resizebox{\hsize}{!}{\includegraphics{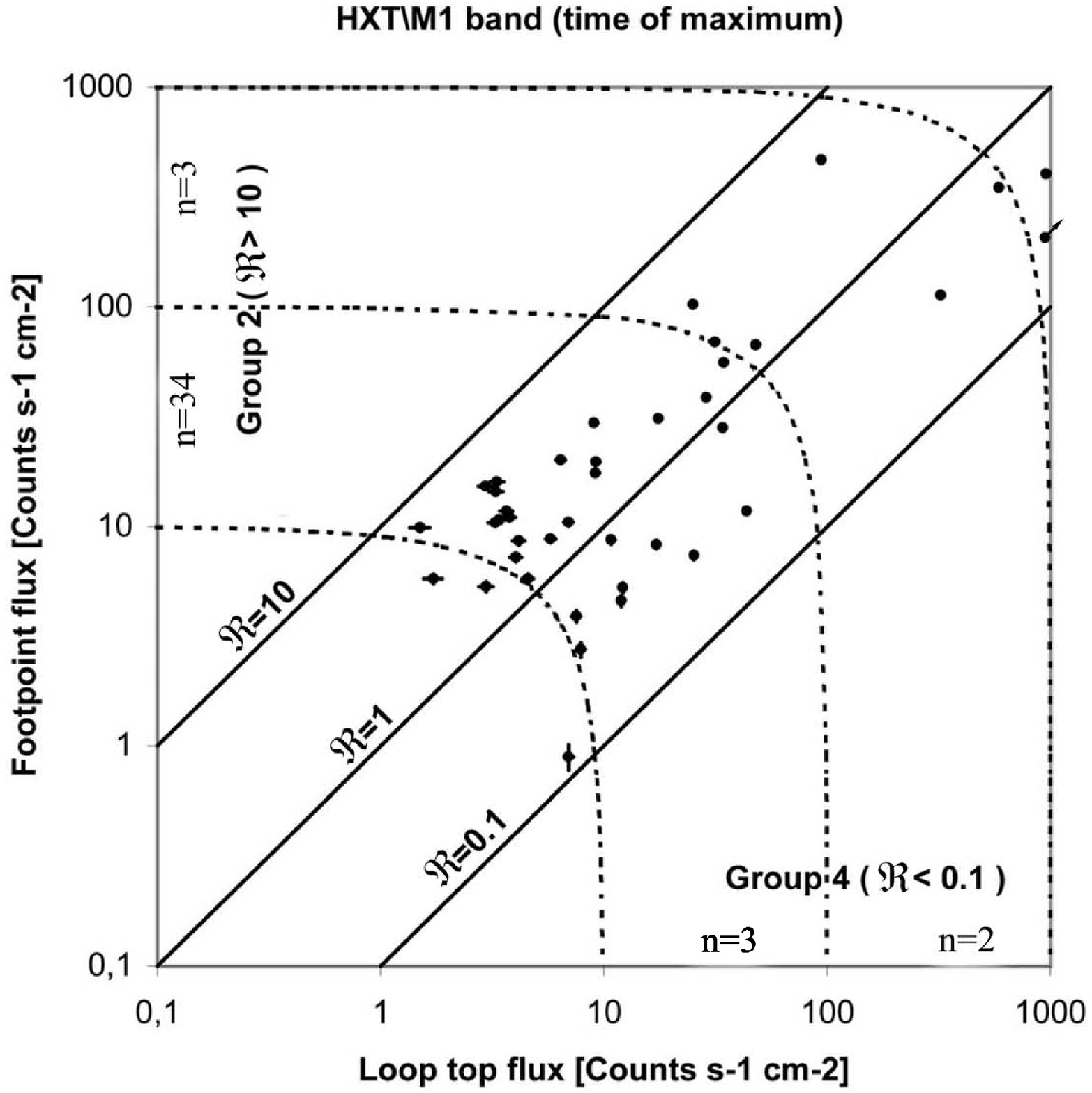}} \hfill
\caption{Counts from the footpoints versus loop-top counts in the
HXT/M1 energy band for 40 flares of group 3. The diagonals represent
lines of constant ratio $\Re$. The lines $\Re = 10$ and $\Re = 0.1$
establish detection thresholds of the instrument. Above the line
$\Re = 10$ we find the group 2 (37 events), below the line $\Re =
0.1$, the group 4 (5 events). The dotted curves show lines of the
constant total flux; the lowest one marks the event selection
threshold.} \label{lt-fp}
\end{figure*}

In Fig.~\ref{lt-fp} we present a plot of the count rates of the
footpoint source versus the count rates of the loop-top source for
events from group 3. In the case of multiple-source events of the
same class, coronal or chromospheric,  we added their fluxes. It
means that in this plot one flare is represented by one point. Error
bars in Fig.~\ref{lt-fp} are statistical uncertainties, thus the
actual uncertainties, including the possibility of contamination of
one source by neighboring ones or an error of the reconstruction
method, should be larger. The diagonals in Fig.~\ref{lt-fp}
represent the line of equal fluxes ($\Re = 1$) and detection
thresholds arising from the finite dynamic range of the HXT. In this
plot events from group 2 are located above the upper diagonal ($\Re
> 10$) and events from group 4 below the lower one ($\Re <
0.1$).

The primary trend seen in Fig.~\ref{lt-fp} is an obvious correlation
between fluxes of both classes of hard X--ray sources over three
decades of flux. This trend mimics a `corridor' between the
diagonals $\Re = 10$ and $\Re = 0.1$ that constitutes an area where
the HXT provides a reasonable detection of sources. Within the
corridor a concentration of events rises as fluxes fall. An empty
area in the lower left corner of the corridor is a truncation effect
introduced by our selection criteria.

The second striking feature in Fig.~\ref{lt-fp} is an increase in
flares for which footpoint sources are brighter than loop-top ones.
Going from the left upper corner to the right bottom one in this
figure we observe a continuous decrease in flare frequency: for 37
events $\Re$ is greater than 10, for 26, $\Re$ is between 10 and 1,
for 14 events, between 1 and 0.1, and for 5 events, below 0.1. This
decrease is independent of flux, therefore we conclude that this is
not an artifact and must be intrinsic to the flare process, another
than a global amount of released energy. Similar results were
obtained by Petrosian et al. (\cite{petrosian2}), however the larger
number of events in our sample allows us to discuss the different
values of the parameter $\Re$ in solar flares.

Petrosian \& Donaghy (\cite{petrosian}) calculated spatial
distributions of hard X--ray emission from flare loops assuming
different values of parameters such loop length, field geometry of
the loop, distribution of non-thermal electrons or plasma density
within the loop. They found that in many cases the distribution is
strongly dominated by footpoint sources because the emission
mechanism is the most efficient there. However, they also obtained
that an enhancement of emission from the loop-top is possible if at
least one of the following conditions is fulfilled:
\begin{enumerate}
\item A pancake-type pitch-angle distribution of the accelerated electrons.
\item Scattering by plasma turbulence at the acceleration site.
\item Converging field geometry of the flare loop.
\item Strong increase of the column depth along the flare loop due to
chromospheric evaporation.
\end{enumerate}
The first two conditions refer to the acceleration process of
non-thermal electrons in solar flares, the last two refer to the
propagation process of non-thermal electrons within the flare loop.
Generally, it is difficult to distinguish the influence of
particular conditions with a reasonable confidence for individual
events. However, we can estimate the general importance of
converging field geometry and column depth along the flare loop for
the ratio $\Re$ for a large sample of events.

The field convergence in the flare loop increases the pitch angle of
the electrons as they descend. As a consequence, part of electrons
can be reflected back by a magnetic mirror up to the loop-top before
the transition region. Thus, the higher magnetic mirror ratio $M_r =
B_{foot}/B_{top}$, the greater the trapping efficiency:
\begin{equation}
M_r = B_{foot}/B_{top} = 1/{\sin}^2{\alpha}_c
\end{equation}
where ${\alpha}_c$ is the critical value of pitch-angle (Aschwanden
\cite{aschwssr}). This means that all electrons having a pitch angle
greater than ${\alpha}_c$ remain in the trap. Thus, an increase of
the magnetic mirror ratio $M_r$ should decrease the hard X--ray flux
ratio $\Re$ due to a decrease of electrons reaching the footpoints
and an increase of electrons remaining in the loop-top.

For limb flares we cannot measure magnetic field strength directly.
However, there is well-known relation between this parameter and
height. Aschwanden et al. (\cite{aschwb-h}) analyzed the
three-dimensional coordinates of 30 loops from the
Extreme-Ultraviolet Imaging Telescope (EIT) 171 \AA\ image on 1996
August 30. They estimated the height dependence of the magnetic
field $B(h)$ of the 30 potential field lines closest to the analyzed
EIT loops and found that it can be approximated by a dipole model:
\begin{equation}
B(h) = B_{foot} {(1 + \frac{h}{h_D})}^{-3},
\end{equation}
where $h_D = 75$ Mm is the mean dipole depth. We can rewrite this
formula using the magnetic mirror ratio, $M_r$, definition:
\begin{equation}
M_r = {(1 + \frac{h}{h_D})}^3.
\end{equation}

From Eqs. (2) and (4) we see that under the assumption of an
isotropic pitch-angle distribution the height increase of the flare
loop from 10 Mm to 20 Mm decreases the number of electrons reaching
the footpoints by about 20\% and increases the number of electrons
remaining in the loop-top by about 33\%. Therefore, we expect that a
height dependence of the flux ratio of hard X--ray sources $\Re(h)$
should exist where $\Re$ drops as $h$ increases. This relation is
independent of the energy of electrons.

The mean free-path of electrons $l_f$ in a volume having electron
density $n_e$ can be estimated according to (Tsuneta et al.
\cite{tsuneta_f}):
\begin{equation}
l_f = 8.3 \times 10^{17} E^2 / n_e,
\end{equation}
where $E$ is the electron energy in keV and $n_e$ is counted per
cm$^{-3}$. Thus, in conditions typical of the pre-flare phase
($n_e\,{\sim}\,10^9{\div}10^{10}$ cm$^{-3}$), $l_f$ is greater than
the loop semi-length. This means that almost all electrons
accelerated at the top of the flare loop easily reach the transition
region producing hard X--ray footpoint sources. An increase of
electron density within the flare loop due to chromospheric
evaporation limits the number of electrons that deposit their energy
in the footpoints. As a consequence, the flux ratio of hard X--ray
sources $\Re(h)$ should decrease as the column depth, $N = n_eL$,
measured along the loop of the semi-length $L$, increases.

\begin{figure*}
\resizebox{\hsize}{!}{\includegraphics{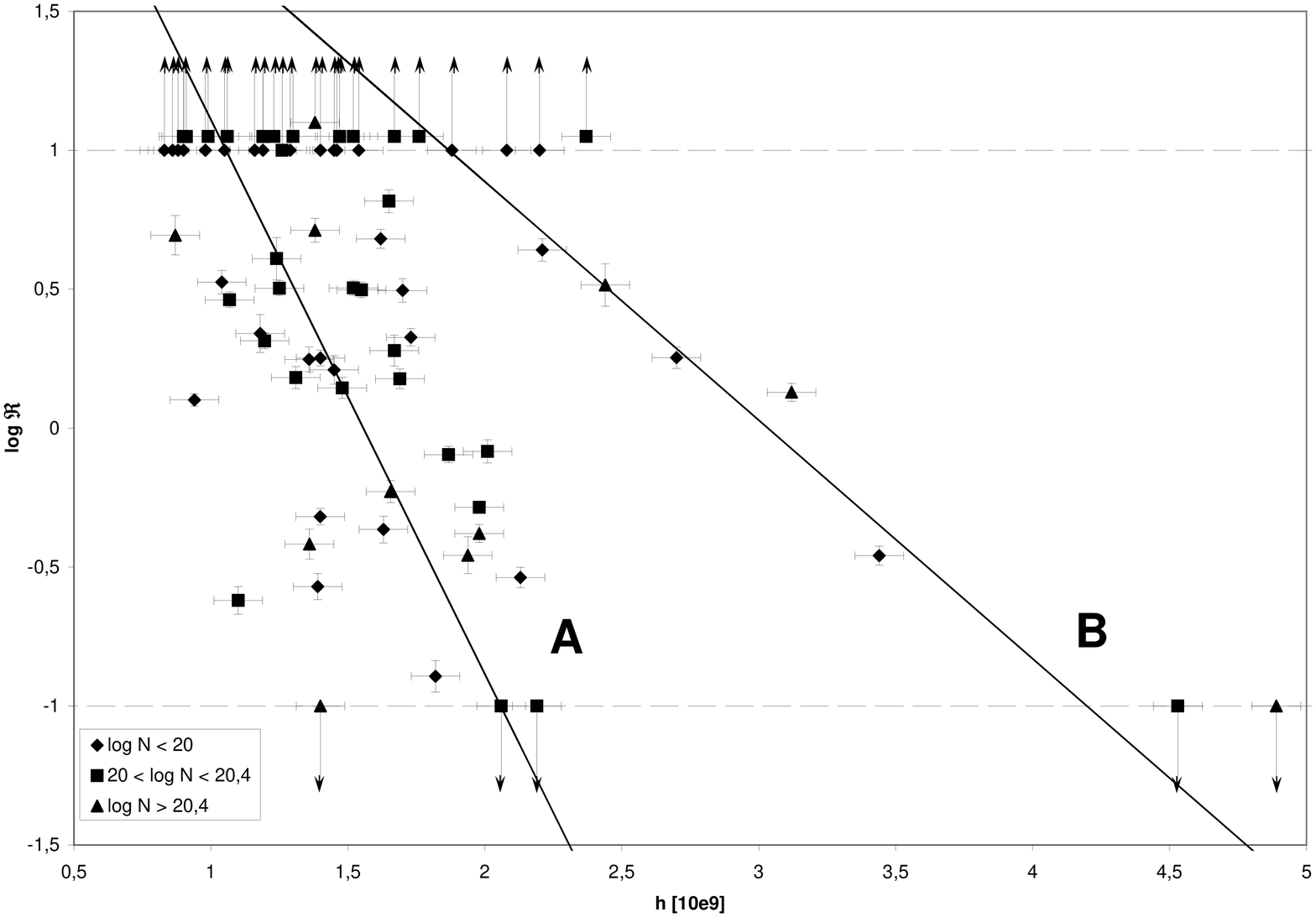}} \hfill
\caption{The ratio $\Re$ (on a logarithmic scale) versus the height
of the flare loop for 82 events. For group 2 and 4 only the lower
and upper approximation of $\Re$ is known, respectively. Different
symbols denote different values of the column depth along the flare
loop. A linear fit to the points from two branches is described by
Eqs. (6) and (7).} \label{r-h}
\end{figure*}

Such relatively simply relationships $\Re(h)$ and $\Re(N)$ can
become more complex if anomalous electron scattering occurs (Melrose
\cite{melrose}). This is caused by MHD turbulence which can be
excited in the coronal segment of the flare loop by anisotropies
occurring in the distribution of accelerated particles. Electrons
with such a distribution can excite whistler waves. The MHD
pitch-angle scattering is more effective than the Coulomb scattering
and it pushes electrons with large pitch angles into the small
pitch-angle regime. Thus, the anomalous scattering contributes to
the footpoint yield electrons which would otherwise be mirrored high
in the corona. The importance of this effect has been proved for
energetic ions (Hua et al. \cite {hua}) and for relativistic
electrons (Miller \& Ramaty \cite{miller}). Recently, Karlick\'y
(\cite{karl2}) has shown that anomalous scattering can be important
also for superthermal electrons.

To determine the relationships $\Re(h)$ and $\Re(N)$ we estimated
values of $h$, $L$, and $n_e$ for flares from Table~\ref{113} using
SXT images. The height, $h$, was directly measured in the Be119
images with a constant error of half of the SXT pixel i.e. about
0.9~Mm. Then, it was corrected for the projection effect by using
flare heliocentric coordinates (column (4) in Table~\ref{113}). We
assumed a perpendicular orientation of loops relative to the solar
surface. To verify our values we also measured the separation
between footpoints and assuming circular loops we obtained an
independent estimation of the height. Differences in obtained values
of $h$ are considered as a measure of their uncertainty. Having $h$
we calculated the semi-length $L$ of the loop, $L=0.5{\pi}h$,
assuming circular loops.

To estimate the electron density within the flare loop we chose a
pair of SXT images taken with the filters Be119 and Al12 closest in
time to the moment of peak maximum in the M1 band. We calculated
values of temperature and emission measure averaged over the whole
flare loop. To estimate the emitting volume we assumed the thickness
of flare loop to be equal to its diameter ${\pm}0.5$ SXT pixel.
Having the emission measure and volume we calculated the value of
$n_e$, thus the column depth $N$. Obtained values of $h$ and $N$ are
given in Table~\ref{113} in column (10) and (11).

For 21 events we could not calculate electron density due to a lack
of useable SXT images for the maximum in the M1 band. As a
substitute we adapted records from the {\sl Geostationary
Environmental Operational Satellites (GOES)}. We used normalization
coefficients for different satellites as well as recent coefficients
for a polynomial approximation to the GOES temperature and emission
measure response (White et al. \cite{goes}). The obtained values of
the column depth (see column (12) in Table~\ref{113}) are
systematically higher than those obtained by the SXT (by a factor 4,
on average).

In Fig.~\ref{r-h} we present the relation between the flux ratio
$\Re$ of hard X--ray sources on a logarithmic scale and the height
for 40 flares for which we obtained values of $\Re$ of between 10
and 0.1. The points are concentrated along two branches which can be
described by the following formulae:
\begin{equation}
\log{\Re} = (-2.05 \pm 0.8) \times h + (3.2 \pm 0.6) \quad\mbox{for
35 points}\quad
\end{equation}
and
\begin{equation}
\log{\Re} = (-0.83 \pm 0.15) \times h + (2.5 \pm 0.4) \quad\mbox{for
5 points}\quad
\end{equation}
In this figure we also plot the points that represent events of
group 2 and 4 for which we know only a lower and upper approximation
of $\Re$, respectively. Almost all points from these two groups are
located along the same branches described with Eqs.~(6) and (7). In
column (13) in Table~\ref{113} we divide the events into branches.
For 11 events from group 2 the branches are located too close to
decide about the membership.

\begin{figure*}
\resizebox{\hsize}{!}{\includegraphics{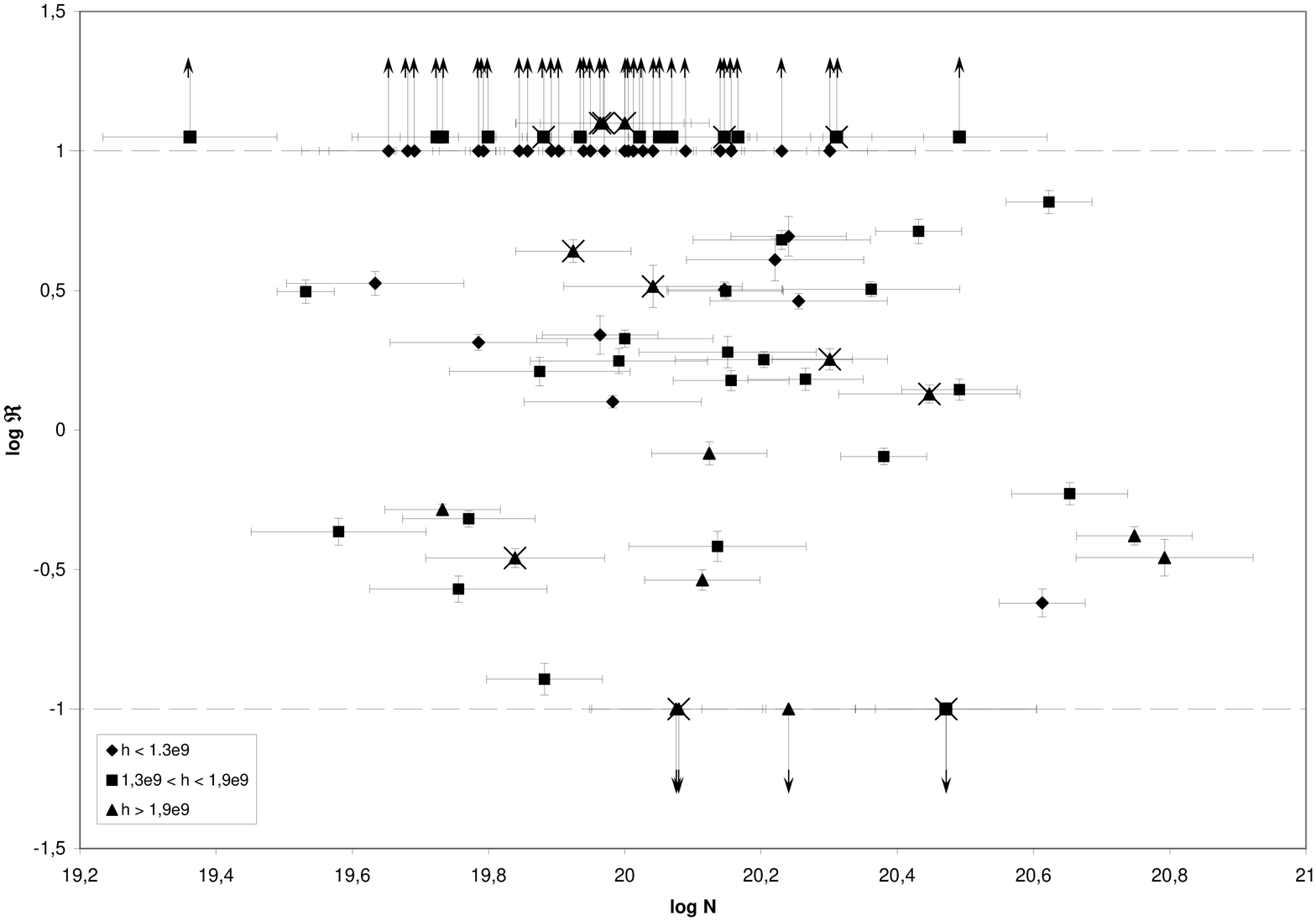}} \hfill
\caption{The ratio $\Re$ versus the column depth along the flare
loop (both on a logarithmic scale) for 82 events. For group 2 and 4
only the lower and upper approximation of $\Re$ is known,
respectively. Different symbols denote different values of the
height of the flare loop. Events from the branch B in Fig.~\ref{r-h}
are marked by crosses.} \label{r-n}
\end{figure*}

The scatter seen in Fig.~\ref{r-h} can be caused by different
statistical as well as systematic factors. Unfortunately, we have no
observational criteria that allow us to study the influence of the
acceleration process of non-thermal electrons in solar flares on the
ratio $\Re$. To separate the influence of the column depth on the
height dependence of the ratio $\Re$ we introduce three symbols:
diamonds, boxes, and triangles, which refer to events with lower
($<1.0{\times}10^{20}$ cm$^{-2}$), intermediate, and higher
($>2.4{\times}10^{20}$ cm$^{-2}$) column depth. In this figure we
used the more complete set of data including the column depths
estimated from GOES data. As we see, different values of the column
depth are not responsible for the separation between the two
branches in Fig.~\ref{r-h} because different symbols are present in
both branches.

In Fig.~\ref{r-n} we show the relation between the flux ratio $\Re$
of hard X--ray sources and the column depth, both on a logarithmic
scale, for the 40 flares for which we obtained values of $\Re$
between 10 and 0.1 and estimated the column depth. In this figure we
also plot the points that represent events of group 2 and 4 for
which we know only the lower and upper approximation of $\Re$,
respectively. We used the more complete set of data including the
column depth estimated from GOES data. No correlation is seen, even
if we plot SXT data instead of GOES ones.

To determine the influence of the height on the column-depth
dependence of the ratio $\Re$ we introduce three symbols: diamonds,
boxes, and triangles which refer to events with lower ($<$13~Mm),
intermediate, and greater ($>$19~Mm) height. The different symbols
in Fig.~\ref{r-n} are distributed almost uniformly within the range
of values of $N$ which suggests that the influence of height is not
responsible for the lack of dependence between the ratio $\Re$ and
the column depth.

\section{Discussion}
\label{disc}

We expected that both $\Re(h)$ and $\Re(N)$ could be detected using
{\sl Yohkoh} data despite the complex interaction between many
parameters and the relatively high values of their uncertainties.
Our analysis confirms the existence of the $\Re(h)$ relationship
only.

\subsection{$\Re(h)$ dependence}
\label{R-h}

We found that the ratio $\Re$ decreases as the height of the flare
loop rises, however, this relationship is complex with two branches:
a left, steeper, and a right, flatter. We excluded different values
of the column depth as the cause of the membership of a particular
branch. We do not think that the acceleration process of non-thermal
electrons decides into which branch the event falls. If so, then all
points from the branch B in Fig.~\ref{r-h} should have a particular
distribution of electrons which makes propagation into the
footpoints more efficient than for other events. Thus, the points
from the branch B in Fig.~\ref{r-h} should also show higher values
of the ratio $\Re$ for similar values of the column depth in
Fig.~\ref{r-n}. These points are marked by crosses and as we see,
this is not the case.

We suggest that events from different branches in Fig.~\ref{r-h}
occurred in magnetic loops having different converging field
geometries. Thus, events from branch A occurred in more converged
loops than the events from branch B. This is confirmed by the
different inclination of the branches. For the branch A an increase
of the height of about 9.5~Mm, on average, transforms the flux ratio
$\Re$ from the value of 10 into the value of 0.1, i.e. it makes the
event of group 4 instead of the event of group 2. To do the same for
the branch B an increase of height of about 24~Mm is needed.

We used our data to estimate the frequency of occurrence of
more-converged and less-converged loops in the solar corona. We have
classified 58 events into branch A and 13 events into branch B (see
column (13) in Table~\ref{113}). This gives an 82\% contribution of
the more-converged loops and an 18\% contribution of the
less-converged ones. For group 2 the branches are situated very
close each to other, therefore a safer method is to take into
account only events of group 3 and 4. In this case the contribution
of the more-converged loops is even higher (38 events of 45 i.e.
84\%) and the contribution of the less-converged ones is 16\% (7
events of 45).

\begin{figure}
\resizebox{\hsize}{!}{\includegraphics{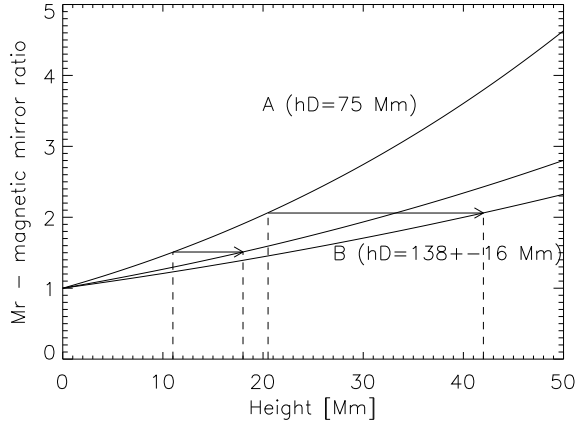}} \hfill
\caption{Changes of the magnetic mirror ratio against the height of
the magnetic loop calculated with formula (4). Different values of
the mean dipole depth, $h_D$, reflect different inclinations of the
branches A and B in Fig.~\ref{r-h}.} \label{mmr}
\end{figure}

To describe the convergence of the investigated magnetic loops more
quantitatively we used formula (4) and assumed {\sl a priori} that
it works for more frequent events from branch A (Fig.~\ref{mmr}). We
calculated the magnetic mirror ratio, $M_r$, for terminal heights,
11 and 20.5~Mm, establishing the group 3 for branch A. The obtained
values of $M_r$ are 1.51 and 2.06, respectively. This means that an
increase of the magnetic mirror ratio by a factor 1.36 completely
changes the spatial distribution of hard X--ray emission in solar
flares. For $M_r \le 1.51$ the population of electrons reaching the
footpoints is so large that the HXT can detect only footpoint
sources. For $M_r \ge 2.06$ the population of electrons trapped
around the loop-top is so large that the HXT can detect only
loop-top sources.

If our explanation is consistent, the magnetic loops from branch B
should reach the same values of $M_r$, 1.51 and 2.06, for greater
terminal heights, 18 and 42~Mm, respectively. To fulfil this
condition the mean dipole depth, $h_D$, in formula (4) should
increase to 138$\pm$16~Mm i.e. by a factor 1.8 (Fig.~\ref{mmr}).

Our results are consistent with Aschwanden et al.
(\cite{aschw-batse}) who estimated the magnetic mirror ratio for 46
solar flares recorded by the Burst and Transient Source Experiment
(BATSE) onboard the {\sl Compton Gamma Ray Observatory (CGRO)}. For
a comparison see Table~\ref{aschw1}.

The observed width variation of coronal loops has been extensively
investigated due to its ability to test magnetic field models in the
solar corona. Many authors have measured the ratio of the loop
diameter at the top to that at the footpoint, the so-called
expansion factor, for selected magnetic loops observed by different
instruments: {\sl Yohkoh}/SXT, {\sl SoHO}/EIT, {\sl TRACE} (Klimchuk
et al. \cite{klim1}, Aschwanden et al. \cite{aschwb-h}, Klimchuk
\cite{klim2}, Watko \& Klimchuk \cite{klim3}, Khan et al.
\cite{khan}, L\'opez Fuentes et al. \cite{lopez}). Different sets of
non-flare and post-flare loops gave average values of the expansion
factor between 1.0 and 1.3. The authors did not report any tendency
towards loop differentiation for separate values of the expansion
factor. Moreover, Watko \& Klimchuk (\cite{klim3}) did not find any
correlation between the expansion factor and loop length.

Our results are different. First, we found the relation $\Re(h)$
which we interpret as a consequence of a correlation between the
expansion factor and the loop length. Second, we identified two
kinds of converging field geometry (the branches A and B in
Fig.~\ref{r-h}). Third, many of the investigated events suggest an
expansion factor above the upper limit of previous investigations.
However, we developed a new method to estimate the expansion factor
of loops in which non-thermal electron beams are used as a
diagnostic tool. This method includes the convergence of loops at
the entrance into the chromosphere in a better way than a direct
measure of the loop diameter ratio because of the limited
sensitivity of SXR and EUV filters due to the temperature decrease.

\begin{table}
\caption[ ]{Comparison between estimations of the magnetic mirror
ratio for flares investigated by Aschwanden et al.
(\cite{aschw-batse}) and in this paper.}
\begin{flushleft}
\label{aschw1}
\begin{tabular}{ccc}
\hline
 & \multicolumn{2}{c}{Number of events} \\
 \cline{2-3}
 Magnetic mirror  & Aschwanden et al.
 &  This paper \\
ratio $M_r$ & (\cite{aschw-batse}) & \\
 \hline
$\le$1.5 & 21 (45.6\%) & 37 (45.1\%) \\
1.5--2.1 & 17 (37.0\%) & 40 (48.8\%) \\
$\ge$2.1 & 8 (17.4\%) & 5 (6.1\%) \\
\hline
Total & 46 (100\%) & 82 (100\%) \\
\end{tabular}
\end{flushleft}
\end{table}

Conventional magnetic field models, including a potential or a
force-free extrapolation of photospheric magnetograms, predict that
the expansion factor should be much larger than those obtained from
observations (Aschwanden \cite{aschwpodr}). To explain the
discrepancy between the observed and expected expansion it has been
postulated that magnetic flux tubes that correspond to observed
plasma loops are current-carrying and a helically twisted (Klimchuk
et al. \cite{klim4}). Such flux tubes show a reduced expansion
factor due to the magnetic tension associated with the azimuthal
field component that is introduced by the twist.

To understand what differentiates the investigated flare loops in
Fig.~\ref{r-h}, we applied a interpretation of helically twisted
flux tubes (Klimchuk et al. \cite{klim4}). In this picture the
more-converged loops from branch A in Fig.~\ref{r-h} correspond to
flux tubes that are less twisted, and the less converged loops (the
branch B in Fig.~\ref{r-h}) correspond to flux tubes that are more
twisted.

To verify this hypothesis we considered three observables which may
depend on the helical twist of the flux tube in which the flare
occurred. They are: (1) temperature of the flare, (2) magnetic
complexity of the active region, and (3) the connection with Coronal
Mass Ejections (CMEs). However, none have shown any systematic
difference between events from branch A and B. A lack of a
difference probably means that the connection between the
observables and the helical twist of the flux tube is too loose and
is missed due to observational uncertainties. A further search for a
more convenient observable representing the helical twist is needed.

When discussing the possibility of the occurrence of two types of
magnetic loops, the more converged and less converged, the same loop
can show a different convergence in its footpoints. Such a solution
was proposed by Sakao (\cite{sakao}) to explain the asymmetry of
footpoint hard X--ray emission sources in five disc flares. He found
that usually the brighter source is co-spatial with an area where
the magnetic field is fainter. He concluded that due to a lower
convergence in this footpoint, non-thermal electrons precipitate
deeper where hard X--ray photons are produced more efficiently.
Based on microwave images, Kundu et al. (\cite{kundu}) confirmed
this for two other flares. In this case the brighter footpoint
source was co-spatial with an area of a stronger magnetic field due
to the dependence of microwave emission on magnetic field intensity.

If Sakao's scheme is correct then the stronger asymmetry of
convergence in the flare loop causes the stronger asymmetry of hard
X--ray emission. As a consequence, the flux ratio $\Re$ depends more
strongly on properties of the brighter, less-converged footpoint.
Thus, the membership of events of the branch B in Fig.~\ref{r-h}
might be caused by the magnetic configuration in which only the one
footpoint is less converged.

\begin{table}
\caption[ ]{Values of the asymmetry ratio $a$ for events from branch
A and B.}
\begin{flushleft}
\label{asymm}
\begin{tabular}{ccc}
\hline
Branch & $a \le 0.67$ & $a > 0.67$ \\
 \hline
 & & \\
A (55 events) & 22 (40\%) & 33 (60\%) \\
 & & \\
B (11 events) & 4 (36\%) & 7 (64\%) \\
 & & \\
 \hline
Total (66 events) & 26 (39\%) & 40 (61\%) \\

\end{tabular}
\end{flushleft}
\end{table}

To verify this possibility we compared the asymmetry of footpoint
sources of events from branch A to those from branch B for events
from groups 2 and 3. Following Aschwanden et al.
(\cite{aschw-asymm}) we defined the asymmetry ratio $a$ of the
footpoint hard X--ray fluxes $F_{FP1}$ and $F_{FP2}$:
\begin{equation}
a = \frac{F_{FP2}}{(F_{FP1} + F_{FP2})}, \quad\mbox{where}\quad
F_{FP1} < F_{FP2}.
\end{equation}
With this definition, the asymmetry ratio varies from $a=0.5$ for
symmetric footpoints to $a=1$ for the one-sided footpoint. We found
that about 40\% of events show small values of the asymmetry ratio
($a \le 0.67$) and about 60\% of events show large values of the
asymmetry ratio ($a > 0.67$). The contribution of symmetric and
asymmetric events in both branches is almost the same (see
Table~\ref{asymm}). Thus, asymmetric convergence in flare loop
footpoints can influence the location of events in the $\Re$--$h$
diagram, however, it does not explain the division into two
branches.

Goff et al. (\cite{goff}) have shown that a different brightness in
flare footpoint hard X--ray sources also can be caused by a
non-central location of the acceleration site in the magnetic loop.
For many events we could not precisely localize the acceleration
site, therefore we did not discuss this possibility.

Employing the effect of anomalous electron scattering in the flare
loop it is possible to give an alternative explanation for presence
of two branches in Fig.~\ref{r-h}: all investigated events show a
similar convergence but in the case of flares from branch B the
anomalous electron scattering operates in an opposite way to events
in branch A. This effect decreases values of the pitch angle of
electrons, therefore for the same height we should observe a higher
value of the ratio $\Re$. Unfortunately, the HXT has a poor energy
resolution so we could not verify the status of the electron
scattering.

\subsection{$\Re(N)$ dependence}
\label{R-N}

Opposite to our expectation we have not found any correlation
between the flux ratio $\Re$ and the column depth $N$ along the
flare loop (see Fig.~\ref{r-n}). A reason for this may be the time
interval chosen for analysis --- the maximum in the HXT/M1 energy
band. It may be too early for efficient operating of the
chromospheric evaporation. This explanation is supported by the fact
that in Fig.~\ref{r-n} the values of $\Re$ less than 1 do not occur
for short loops ($h<$\,13~Mm) with the only exception of the strong
flare of August 18, 1998 (event No.~49).

\begin{figure}
\resizebox{\hsize}{!}{\includegraphics{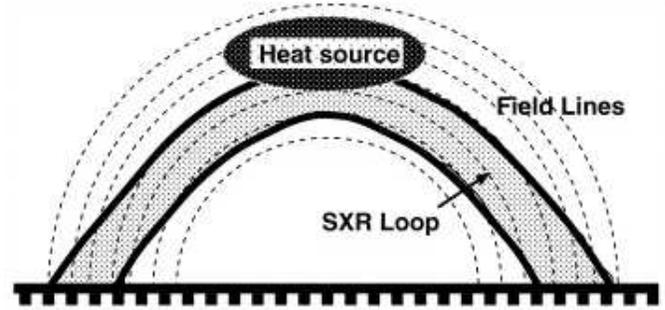}} \hfill
\caption{Reproduced from Hori et al. (\cite{hori}). Field lines
connecting a heat source and footpoints extend outside the bright
loop seen in soft X-rays which is an apparent product of successive
energy release and inertia of the chromospheric evaporation.}
\label{hori}
\end{figure}

The lack of correlation in Fig.~\ref{r-n} can be explained also by
the scheme presented in Fig.~\ref{hori}. This figure is reproduced
from Hori et al. (\cite{hori}). As we see, field lines connecting a
heat source and footpoints extend beyond the bright loop seen in
soft X-rays. The impression that we observe a single bright loop was
caused by successive energy release from the innermost loop to the
outermost and the inertia of chromospheric evaporation. If this
scheme is correct we should rather measure the column depth outside
the bright legs seen in SXT images which apparently connect the
loop-top and footpoint hard X--ray sources. The scheme from
Fig.~\ref{hori} does not need to work for each of the investigated
events to argue against a correlation between the flux ratio and the
column depth. Anomalous electron scattering can be an additional
cause that makes the correlation between $\Re$ and $N$ unclear.

\section{Conclusions}
\label{concl}

We investigated the spatial distribution of hard X--ray emission in
solar flares. We concentrated on the relationship between the
loop-top and footpoint emission sources. In comparison to previous
studies (Masuda \cite{masuda}, Petrosian et al. \cite{petrosian2})
the list of investigated flares is more complete (117 events instead
of 10 or 18, respectively) due to the mission-long {\sl Yohkoh}/HXT
database that was included and the more liberal selection criteria
that were used.

We investigated the dependence of the flux ratio $\Re$ of the
footpoint and loop-top sources on other observational parameters
like the flare loop height $h$ and the column depth $N$ along the
flare loop. We found a clear correlation between $\Re$ and $h$
(Fig.~\ref{r-h}) which we interpret as an effect of converging field
geometry. Since this correlation is stronger than other parameters
and observational uncertainties we conclude that the influence of
magnetic convergence on the spatial distribution of hard X--ray
emission in solar flares is the strongest, at least under conditions
defined in this investigation i.e during the early phase of the
flare evolution. This is confirmed, to some extent, by the lack of
any correlation between the flux ratio $\Re$ and the column depth
$N$ (Fig.~\ref{r-n}). However, this can be also caused by a more
complex configuration of investigated events than is seen in soft
X--rays (see Fig.~\ref{hori}).

Using the formula from Aschwanden et al. (\cite{aschwb-h}) we
developed a new method of calculating the magnetic mirror ratio in
solar flare loops using non-thermal electron beams as a diagnostic
tool for magnetic convergence. The obtained values of the magnetic
mirror ratio are consistent with previous results and suggest a
nonpotential configuration strongly distorted by the presence of
longitudinal currents.

The two branches seen in the $\Re-h$ diagram suggest that in the
solar corona two kinds of magnetic loops can exist: a more converged
kind that is more frequent (above 80\%) and a less converged kind
that is less frequent (below 20\%). In a dipole approximation the
mean dipole depth $h_D$ differs between the branches by a factor of
1.8, on average. Despite the observed dichotomy in solar flare loops
into more converged and less converged cases, we have no independent
proof confirming such a division. Therefore, we cannot conclude,
whether the division is real or random. Finally, what is a role of
the anomalous electron scattering? Further studies are needed
including more and better observed flares to answer these questions.
For this, the {\sl Reuven Ramaty High-Energy Solar Spectroscopic
Imager (RHESSI)} is ideal due to its dynamic range and energy
resolution better than the HXT. This work is in progress.

We hope that further study will allow us to understand also the
concentration in the rarely-occupied branch B in Fig.~\ref{r-h} of
many known and extensively investigated flares e.g. of January 13,
1992 (Masuda et al. \cite{masuda-nat}), of February 17, 1992
(Doschek et al. \cite{doschek}), of June 28, 1992 (Tomczak
\cite{tomczak-pajak}), and of October 4, 1992 (Masuda et al.
\cite{masuda-pasj}).

\begin{acknowledgements}

The {\sl Yohkoh} satellite is a project of the Institute of Space
and Astronautical Science of Japan. The authors are grateful to the
referee, Dr. Marian Karlick\'y, for his valuable remarks which
helped to improve the paper. This work was supported by a Polish
Ministry of Science and High Education grant.

\end{acknowledgements}

\onecolumn \setlongtables
\begin{small}
\begin{longtable}{rccccrlcccrrcr}
\caption{List of selected flares ($^{\rm a}$ Values in parenthesis
are estimated from SXT images; $^{\rm b}$ M -- Masuda \cite{masuda},
P -- Petrosian et al. \cite{petrosian2}).} \label{113}
\\
\hline No. & Date & GOES & Disk & NOAA & \multicolumn{1}{c}{M1} &
\multicolumn{1}{c}{Time} & Group & $\Re$ & $h$ &
\multicolumn{2}{c}{$N$ [10$^{20}$~cm$^{-2}$]}& Branch & Rem.$^{\rm
b}$
\\ \cline{11-12}
 & & class & position$^{\rm a}$ & AR & \multicolumn{1}{c}{cts.} & \multicolumn{1}{c}{interval} & & & [Mm] & SXT & GOES & [Fig.~\ref{r-h}] & \\
  (1) & (2) & (3) & (4) & (5) & (6) & \multicolumn{1}{c}{(7)} & (8) & (9) & (10) & (11) & (12) & (13) & (14) \\
\hline
\endfirsthead
\caption{cont.}\\
\hline
 (1) & (2) & (3) & (4) & (5) & (6) & \multicolumn{1}{c}{(7)} & (8) & (9) & (10) & (11)
 & (12) & (13) & (14) \\
 \hline
 \endhead
   1 & 91/11/09 & M1.4 & S14W69 & 6906 &  105 & 20:52:09+2   & 2 &      & 15 & 0.34 & 1.4 \hspace*{0.08cm} & ? &      \\
   2 & 91/11/17 & M1.9 & S12E78 & 6929 &   49 & 18:33:17+2   & 2 &      & 10 & 0.20 & 1.7 \hspace*{0.08cm}  & A &      \\
   3 & 91/12/02 & M3.6 & N16E87 & 6955 &   24 & 04:53:44+4.5 & 3 & 0.48 & 14 & 0.21 & 0.59 & A & M, P \\
   4 & 91/12/03 & X2.2 & N17E72 & 6955 & 1297 & 16:36:30+0.5 & 1 &      & 12 & 0.79 & 2.2 \hspace*{0.08cm}  &   &      \\
   5 & 91/12/07 & C6.4 & N09E72 & 6961 &   17 & 21:52:46+7.5 & 1 &      & 12 & 0.52 & 0.27 &   &      \\
   6 & 91/12/08 & M2.6 & N08E65 & 6961 &   12 & 07:35:00+9   & 1 &      & 15 & 0.61 & 0.92 &   &      \\
   7 & 91/12/09 & M1.1 & S06E88 & 6966 &   13 & 18:53:46+9.5 & 3 &  6.6 \hspace*{0.08cm} & 17 & 0.31 & 4.1 \hspace*{0.08cm}  & A &      \\
   8 & 91/12/15 & C6.0 & S12E76 & 6972 &   14 & 14:21:42+5   & 2 &      & 12 & 0.20 & 0.62 & A & M, P \\
   9 & 91/12/16 & C7.8 & S10E69 & 6972 &   43 & 03:13:06+2.5 & 2 &      & 10 & 0.27 & 0.80 & A &      \\
  10 & 91/12/16 & M1.6 & S12E67 & 6972 &   50 & 06:38:15+2.5 & 2 &      & 9 \hspace*{-0.3cm} & 1.4 \hspace*{0.08cm}  & 1.4 \hspace*{0.08cm}  & A &      \\
  11 & 91/12/18 & M3.5 & S10E88 & 6980 &  106 & 10:27:30+1.5 & 3 &  1.4 \hspace*{0.08cm} & 15 & 0.31 & 3.1 \hspace*{0.08cm}  & A & M, P \\
  12 & 92/01/13 & M2.0 & S16W86 & 6994 &   29 & 17:28:11+3.5 & 2 &      & 24 & 0.11 & 0.93 & B & M, P \\
  13 & 92/02/06 & M7.6 & N05W82 & 7030 &  144 & 03:23:50+1   & 4 &      & 22 & 0.38 & 1.7 \hspace*{0.08cm}  & A & M, P \\
  14 & 92/02/17 & M1.9 & S16W81 & 7050 &   21 & 15:40:54+5   & 2 &      & 17 & 0.16 & 0.63 & B & M, P \\
  15 & 92/02/19 & M3.7 & N04E85 & 7067 &   17 & 03:48:23+6.5 & 3 &  4.4 \hspace*{0.08cm} & 22 & 0.23 & 0.84 & B &      \\
  16 & 92/02/26 & M1.3 & S16W90 & 7073 &   15 & 01:37:58+6.5 & 2 &      & 12 & 0.33 & 2.0 \hspace*{0.08cm}  & A &      \\
  17 & 92/04/01 & M2.3 & S04E86 & 7123 &   38 & 10:13:04+3   & 2 &      &  9 \hspace*{-0.3cm} & 0.64 & 1.0 \hspace*{0.08cm}  & A & M, P \\
  18 & 92/04/19 & C3.9 & N04E85 & 7138 &   16 & 02:11:11+6.5 & 2 &      & 22 & 0.11 & 0.92 & B &      \\
  19 & 92/06/26 & C2.9 & N08W82 & 7205 &   11 & 12:52:15+9.5 & 1 &      &  9 \hspace*{-0.3cm} & 0.26 & 0.97 &   &      \\
  20 & 92/06/28 & M1.6 & N10W89 & 7216 &   17 & 13:56:43+5.5 & 4 &      & 45 &      & 1.1 \hspace*{0.08cm}  & B &      \\
  21 & 92/08/11 & M1.4 &(S15E81)& 7260 &   77 & 22:25:20+1.5 & 1 &      & 14 & 0.19 & 0.42 &   &      \\
  22 & 92/09/09 & M3.1 & S09W71 & 7270 &   32 & 02:10:29+3.5 & 1 &      & 16 & 0.27 & 1.7 \hspace*{0.08cm}  &   &      \\
  23 & 92/09/09 & M1.9 & S11W78 & 7270 &   15 & 17:58:33+6   & 1 &      & 14 & 0.17 & 1.5 \hspace*{0.08cm}  &   &      \\
  24 & 92/10/04 & M2.4 & S05W90 & 7293 &   30 & 22:19:07+3   & 2 &      & 19 & 0.91 & 1.4 \hspace*{0.08cm}  & B & M, P \\
  25 & 92/10/11 & C8.3 & S16W68 & 7301 &   14 & 02:03:36+10.5& 2 &      & 15 & 0.14 & 0.54 & ? &      \\
  26 & 92/11/05 & M2.0 & S17W84 & 7323 &   36 & 06:19:22+3   & 1 &      & 10 & 0.40 & 1.0 \hspace*{0.08cm}  &   & M, P \\
  27 & 92/11/22 & M1.6 & N11E76 & 7348 &   31 & 23:07:49+4   & 2 &      &  9 \hspace*{-0.3cm} & 0.73 & 0.61 & A &      \\
  28 & 92/12/04 & M1.4 & N19W71 & 7352 &   47 & 11:38:32+2   & 2 &      & 15 & 0.31 & 0.86 & ? &      \\
  29 & 93/02/05 & C7.7 & S07E78 & 7420 &   29 & 07:53:38+3.5 & 1 &      & 15 &      & 0.55 &   &      \\
  30 & 93/02/06 & C5.6 & S08E66 & 7420 &   23 & 05:25:56+4.5 & 1 &      &  8 \hspace*{-0.3cm} &      & 0.90 &   &      \\
  31 & 93/02/14 & M2.0 & S22E78 & 7427 &   53 & 12:53:04+3   & 1 &      & 14 & 0.50 & 2.3 \hspace*{0.08cm}  &   &      \\
  32 & 93/02/17 & M5.8 & S07W87 & 7420 &   94 & 10:36:21+1   & 3 &  2.2 \hspace*{0.08cm} & 12 & 0.22 & 0.92 & A & M, P \\
  33 & 93/02/21 & M1.4 & N13E75 & 7433 &   14 & 00:39:35+8.5 & 3 &  2.9 \hspace*{0.08cm} & 11 & 0.60 & 1.8 \hspace*{0.08cm}  & A &      \\
  34 & 93/03/15 & C3.0 & N06W88 & 7411 &   15 & 09:59:44+9   & 2 &      & 12 & 0.11 & 0.92 & A &      \\
  35 & 93/06/11 & C5.7 & S13W80 & 7518 &   12 & 10:14:59+9.5 & 3 & 0.35 & 34 &      & 0.69 & B &      \\
  36 & 93/06/25 & M5.1 &(S09E90)& 7530 &   17 & 03:18:23+6   & 3 &  1.5 \hspace*{0.08cm} & 13 & 0.22 & 1.4 \hspace*{0.08cm}  &
  A &      \\
  37 & 93/09/27 & M1.8 &(N09E89)& 7590 &   41 & 12:08:14+4.5 & 2 &
     &  9 \hspace*{-0.3cm} & 0.62 & 1.1 \hspace*{0.08cm} & A & P \\
  38 & 93/10/09 & M1.1 & N11W78 & 7590 &   11 & 08:08:18+10  & 3 &  3.4 \hspace*{0.08cm} & 10 & 0.25 & 0.43 & A &      \\
  39 & 93/11/30 & C9.2 &(N19E84)& 7618 &   70 & 06:03:31+4   & 2 &
     & 14 & 0.21 & 1.0 \hspace*{0.08cm} & ? & P \\
  40 & 94/01/16 & M6.1 & N09E70 & 7654 &   57 & 23:17:27+2   & 3 & 0.83 & 20 & 0.51 & 1.3 \hspace*{0.08cm}  & A &      \\
  41 & 94/01/17 & C9.3 & N06E65 & 7654 &   12 & 09:14:49+10.5& 2 &      & 15 & 0.16 & 1.2 \hspace*{0.08cm}  & ? &      \\
  42 & 97/08/09 & C8.5 & N19W85 & 8069 &   12 & 16:33:35+10  & 1 &      &  9 \hspace*{-0.3cm} &      & 1.4 \hspace*{0.08cm}  &   &      \\
  43 & 97/09/14 & C2.8 & S23W79 & 8083 &   17 & 02:53:26+6   & 2 &      & 15 &      & 0.23 & ? &      \\
  44 & 97/11/15 & M1.0 & N20E65 & 8108 &   15 & 22:42:04+5.5 & 4 &      & 49 &      & 3.2 \hspace*{0.08cm}  & B &      \\
  45 & 98/05/08 & M3.1 &(S16W90)& 8210 &   48 & 01:57:35+2   & 3 &  1.8 \hspace*{0.08cm} & 14 & 0.19 & 1.6 \hspace*{0.08cm}  & A & P    \\
  46 & 98/05/28 & C8.7 &(N16W89)& 8226 &   12 & 19:02:55+10  & 3 &  2.1 \hspace*{0.08cm} & 12 & 0.15 & 0.61 & A &      \\
  47 & 98/08/14 & M3.1 & S23W74 & 8293 &   65 & 08:25:57+2   & 1 &      & 10 & 1.4 \hspace*{0.08cm}  & 2.2 \hspace*{0.08cm}  &   &      \\
  48 & 98/08/18 & X2.8 &(N34E85)& 8307 &  599 & 08:20:29+0.5 & 4 &
     & 14 & 1.1 \hspace*{0.08cm} & 3.1 \hspace*{0.08cm} & A & P \\
  49 & 98/08/18 & X4.9 & N33E87 & 8307 & 2819 & 22:16:40+0.5 & 3 & 0.24 & 11 & 1.2 \hspace*{0.08cm}  & 4.1 \hspace*{0.08cm}  & A & P    \\
  50 & 98/09/02 & M2.2 &(N18W82)& 8319 &   20 & 17:03:50+5   & 2 &      & 13 & 0.35 & 0.48 & ? &      \\
  51 & 98/09/28 & C9.5 &(S14W90)& 8342 &   26 & 12:00:22+4.5 & 2 &      &  9 \hspace*{-0.3cm} & 0.20 & 1.4  \hspace*{0.08cm} & A &      \\
  52 & 98/09/28 & C6.8 &(S14W90)& 8342 &   21 & 16:08:14+5.5 & 2 &      &  9 \hspace*{-0.3cm} & 0.17 & 0.45 & A &      \\
  53 & 98/11/09 & C4.9 & N22W70 & 8375 &   21 & 21:12:21+5.5 & 3 &  4.8 \hspace*{0.08cm} & 16 & 0.20 & 1.7 \hspace*{0.08cm}  & A &      \\
  54 & 98/11/10 & C7.9 & N20W76 & 8375 &   51 & 00:12:25+2   & 1 &      & 11 & 0.64 & 1.3 \hspace*{0.08cm}  &   &      \\
  55 & 98/11/10 & C3.3 & N21W76 & 8375 &   18 & 06:52:25+6   & 1 &      & 12 & 0.77 & 4.0 \hspace*{0.08cm}  &   &      \\
  56 & 98/11/10 & M1.8 & N21W78 & 8375 &   56 & 15:42:44+2   & 1 &      & 15 & 0.23 & 2.0 \hspace*{0.08cm}  &   &      \\
  57 & 98/11/11 & M1.0 & N22W86 & 8375 &   18 & 04:05:54+6   & 3 &  3.2 \hspace*{0.08cm} & 15 & 0.62 & 2.3 \hspace*{0.08cm}  & A &      \\
  58 & 98/11/11 & C3.5 & N22W90 & 8375 &   13 & 07:36:32+10  & 1 &      & 14 & 0.28 & 0.85 &   &      \\
  59 & 98/11/22 & X3.7 & S29W80 & 8384 & 1200 & 06:39:38+0.5 & 3 & 0.42 & 20 & 0.86 & 5.6 \hspace*{0.08cm}  & A &      \\
  60 & 98/11/22 & X2.5 & S29W86 & 8384 &  830 & 16:20:42+0.5 & 3 & 0.59 & 17 & 1.6 \hspace*{0.08cm}  & 4.5 \hspace*{0.08cm}  & A &      \\
  61 & 98/11/23 & C4.9 &(S30W89)& 8386 &   14 & 05:58:55+9.5 & 3 & 0.51 & 20 & 0.11 & 0.54 & A &      \\
  62 & 98/11/24 & C8.4 & N19E77 & 8395 &   18 & 22:12:33+6.5 & 2 &      & 12 & 0.12 & 1.0 \hspace*{0.08cm}  & A &      \\
  63 & 98/11/25 & C6.4 &(N19E69)& 8395 &   14 & 14:01:01+9.5 & 2 &      &  8 \hspace*{-0.3cm} & 0.30 & 0.87 & A &      \\
  64 & 99/01/14 & M3.0 &(N20E65)& 8460 &   12 & 10:14:42+9.5 & 1 &      & 21 &      & 2.0 \hspace*{0.08cm}  &   &      \\
  65 & 99/06/20 & C5.5 &(N16E90)& 8594 &   18 & 08:36:24+7   & 3 &  5.2 \hspace*{0.08cm} & 14 & 0.10 & 2.7 \hspace*{0.08cm}  & A &      \\
  66 & 99/08/04 & C5.4 & S22W70 & 8645 &   14 & 23:33:52+9   & 1 &      & 12 & 0.27 & 0.87 &   &      \\
  67 & 99/08/20 & C7.8 & S23E66 & 8674 &   21 & 19:22:44+5.5 & 2 &      & 11 & 0.40 & 1.2 \hspace*{0.08cm}  & A &      \\
  68 & 99/09/01 & C6.6 & S28W76 & 8674 &   27 & 18:56:25+4   & 2 &      & 21 & 0.19 & 0.76 & B &      \\
  69 & 99/09/19 & C9.6 & N20W77 & 8699 &   13 & 23:06:39+9   & 3 &  1.8 \hspace*{0.08cm} & 27 & 0.19 & 0.98 & B &      \\
  70 & 99/10/26 & M3.7 &(S15W88)& 8737 &   18 & 21:21:33+7   & 3 & 0.80 & 19 & 0.42 & 2.4 \hspace*{0.08cm}  & A &      \\
  71 & 99/10/27 & M1.0 & S12W89 & 8737 &   12 & 09:09:48+11  & 1 &      & 10 & 0.79 & 1.1 \hspace*{0.08cm}  &   &      \\
  72 & 99/11/07 & C3.1 &(N10E83)& 8759 &   15 & 07:35:22+8   & 3 &  1.5 \hspace*{0.08cm} & 17 & 0.07 & 1.8 \hspace*{0.08cm}  & A &      \\
  73 & 99/11/27 & X1.4 & S15W68 & 8771 &  351 & 12:11:09+0.5 & 1 &      & 13 & 1.6 \hspace*{0.08cm}  & 3.2 \hspace*{0.08cm} &   &      \\
  74 & 99/12/03 & C6.3 & N13E80 & 8788 &   19 & 19:51:22+5.5 & 2 &      &  9 \hspace*{-0.3cm} & 0.54 & 1.0 \hspace*{0.08cm}  & A &      \\
  75 & 99/12/18 & M1.5 & N20E66 & 8806 &  259 & 19:11:51+0.5 & 2 &      & 14 & 0.76 & 3.1 \hspace*{0.08cm}  & ? &      \\
  76 & 00/02/04 & M3.0 & N27E79 & 8858 &  119 & 09:15:09+1   & 3 &  4.1 \hspace*{0.08cm} & 12 & 0.33 & 1.7 \hspace*{0.08cm}  & A &      \\
  77 & 00/02/19 & C4.5 & S19W81 & 8878 &   13 & 06:17:02+10  & 1 &      & 13 & 0.18 & 0.52 &   &      \\
  78 & 00/03/07 & M1.0 &(S16E88)& 8902 &   28 & 19:47:56+3.5 & 1 &      & 10 & 0.23 & 1.8 \hspace*{0.08cm}  &   &      \\
  79 & 00/03/07 & C8.7 & S15E84 & 8902 &  170 & 21:55:45+1   & 2 &      & 10 & 0.09 & 0.48 & A &      \\
  80 & 00/03/27 & C8.4 & S09W69 & 8926 &   37 & 13:59:04+2.5 & 1 &      & 10 & 0.49 & 0.78 &   &      \\
  81 & 00/05/11 & C3.6 & S15E84 & 8998 &   11 & 22:23:17+10.5& 1 &      &  8 \hspace*{-0.3cm} & 0.30 & 0.96 &   &      \\
  82 & 00/05/15 & M1.2 & S20W66 & 8993 &   22 & 18:01:00+6   & 1 &      & 12 & 0.71 & 1.0 \hspace*{0.08cm}  &   &      \\
  83 & 00/05/23 & C4.3 & S22W76 & 8996 &   26 & 17:50:15+4.5 & 1 &      & 11 & 0.31 & 1.0 \hspace*{0.08cm}  &   &      \\
  84 & 00/06/01 & M2.5 & N19E80 & 9026 &   83 & 06:12:02+1.5 & 3 &  1.6 \hspace*{0.08cm} & 15 &      & 0.75 & A &      \\
  85 & 00/07/01 & M1.5 & N11W85 & 9054 &   14 & 23:23:14+9.5 & 3 &  3.2 \hspace*{0.08cm} & 16 & 0.20 & 1.4 \hspace*{0.08cm}  & A &      \\
  86 & 00/07/13 & C9.8 & N17W77 & 9070 &   23 & 07:00:45+5   & 1 &      &  8 \hspace*{-0.3cm} & 0.85 & 2.9 \hspace*{0.08cm}  &   &      \\
  87 & 00/07/13 & M1.5 & N16W84 & 9070 &   42 & 22:03:38+2.5 & 1 &      & 10 & 0.91 & 2.0 \hspace*{0.08cm}  &   &      \\
  88 & 00/07/14 & M1.5 &(N17W83)& 9070 &   66 & 00:43:17+2   & 1 &      & 18 & 0.61 & 4.6 \hspace*{0.08cm}  &   &      \\
  89 & 00/07/26 & M1.3 &(S20W90)& 9087 &   18 & 07:41:57+6.5 & 4 &      & 21 & 0.36 & 1.2 \hspace*{0.08cm}  & A &      \\
  90 & 00/07/27 & M2.4 & N12W71 & 9090 &   65 & 04:08:16+1.5 & 2 &      & 12 & 1.1 \hspace*{0.08cm}  & 1.0 \hspace*{0.08cm}  & A &      \\
  91 & 00/07/27 & M1.5 &(S11W90)& 9091 &   18 & 16:47:57+7   & 3 & 0.43 & 16 &      & 0.38 & A &      \\
  92 & 00/09/30 & X1.2 &(N07W90)& 9169 &  502 & 23:19:29+0.5 & 3 &  4.9 \hspace*{0.08cm} &  9 \hspace*{-0.3cm} & 0.58 & 1.7 \hspace*{0.08cm}  & A &      \\
  93 & 00/10/01 & M2.2 &(N07W90)& 9169 &   52 & 14:00:38+2   & 3 &  1.9 \hspace*{0.08cm} & 17 & 0.34 & 1.4 \hspace*{0.08cm}  & A &      \\
  94 & 00/10/14 & M1.1 & N04W82 & 9182 &   16 & 08:36:39+6   & 3 & 0.38 & 14 & 0.34 & 1.4 \hspace*{0.08cm}  & A &      \\
  95 & 00/10/16 & C7.0 &(N04W90)& 9182 &   26 & 05:42:54+4.5 & 3 &  3.1 \hspace*{0.08cm} & 17 & 0.06 & 0.34 & A &      \\
  96 & 00/10/28 & C9.7 & N15E83 & 9212 &   10 & 07:07:25+10  & 3 & 0.13 & 18 & 0.24 & 0.75 & A &      \\
  97 & 00/12/06 & M1.6 &(S11W66)& 9246 &   12 & 22:25:21+11  & 3 &  1.8 \hspace*{0.08cm} & 14 & 0.62 & 2.0 \hspace*{0.08cm}  & A &      \\
  98 & 01/01/25 & C7.4 & N10E73 & 9325 &   12 & 07:11:20+9   & 2 &      &  9 \hspace*{-0.3cm} & 0.41 & 0.70 & A &      \\
  99 & 01/03/21 & M1.8 &(S06W66)& 9373 &   12 & 02:35:12+10  & 3 &  1.3 \hspace*{0.08cm} &  9 \hspace*{-0.3cm} & 0.59 & 0.96 & A &      \\
 100 & 01/03/21 & C9.8 & S07W70 & 9373 &   34 & 11:25:12+4   & 2 &      & 11 &      & 0.87 & A &      \\
 101 & 01/04/02 & X20  &(N16W70)& 9393 &  394 & 21:36:16+0.5 & 3 & 0.35 & 19 & 2.3 \hspace*{0.08cm}  & 6.2 \hspace*{0.08cm}  & A &      \\
 102 & 01/04/04 & C6.4 &(N15W89)& 9393 &   28 & 03:47:21+4   & 3 &  2.1 \hspace*{0.08cm} & 17 &      & 1.0 \hspace*{0.08cm}  & A &      \\
 103 & 01/04/05 & M3.1 &(N14W90)& 9393 &   65 & 02:07:37+2   & 3 &  1.4 \hspace*{0.08cm} & 31 &      & 2.8 \hspace*{0.08cm}  & B &      \\
 104 & 01/04/05 & M8.4 &(N14W90)& 9393 &   27 & 08:48:59+4   & 2 &      & 18 & 0.68 & 2.0 \hspace*{0.08cm}  & B &      \\
 105 & 01/06/30 & C6.0 & N05E78 & 9562 &   15 & 20:41:07+9.5 & 2 &      & 13 &      & 0.72 & ? &      \\
 106 & 01/08/28 & C6.1 & N15E81 & 9600 &   48 & 02:00:49+2   & 2 &      & 15 &      & 1.1 \hspace*{0.08cm}  & ? &      \\
 107 & 01/09/03 & M2.5 &(S23E88)& 9605 &   18 & 18:23:23+7   & 2 &      & 12 &      & 0.78 & A &      \\
 108 & 01/09/13 & C3.7 & S11E73 & 9616 &   19 & 00:36:50+5.5 & 1 &      & 10 &      & 0.90 &   &      \\
 109 & 01/09/30 & M1.0 & S21W72 & 9628 &   38 & 11:34:26+2   & 3 &  3.3 \hspace*{0.08cm} & 24 &      & 1.1 \hspace*{0.08cm}  & B &      \\
 110 & 01/10/01 & M9.1 &(S20W89)& 9628 &   53 & 05:11:42+2   & 3 & 0.27 & 14 &      & 0.57 & A &      \\
 111 & 01/10/01 & M1.2 &(S17W77)& 9632 &   39 & 23:43:56+2.5 & 1 &      & 12 &      & 1.6 \hspace*{0.08cm}  &   &      \\
 112 & 01/10/02 & C4.7 &(S20W90)& 9632 &   21 & 17:12:16+4   & 2 &      & 15 &      & 0.53 & ? &      \\
 113 & 01/10/29 & M1.3 &(N12W88)& 9673 &   18 & 01:56:32+6   & 3 &  3.2 \hspace*{0.08cm} & 13 & 0.53 & 1.4 \hspace*{0.08cm}  & A &      \\
 114 & 01/11/01 & M1.3 &(S19E86)& 9687 &   28 & 06:50:36+3.5 & 3 & 0.29 & 21 & 0.14 & 1.3 \hspace*{0.08cm}  & A &      \\
 115 & 01/11/06 & C5.6 & S16E75 & 9690 &   35 & 01:31:15+3   & 1 &      &  9 \hspace*{-0.3cm} & 0.40 & 1.4 \hspace*{0.08cm}  &   &      \\
 116 & 01/11/06 & M1.2 & S17E74 & 9690 &   19 & 06:24:16+5   & 1 &      & 19 & 0.23 & 1.4 \hspace*{0.08cm}  &   &      \\
 117 & 01/11/06 & C9.0 & S17E73 & 9690 &   10 & 09:28:16+10  & 1 &      & 10 & 0.71 & 2.3 \hspace*{0.08cm}  &   &      \\
\hline
\end{longtable}
\end{small}
\twocolumn

\end{document}